\documentstyle[emulateapj]{article}
\newcommand{\eg}{e.g.,\ }

\newcommand{\hh}{{\rm H}_2}
\newcommand{\Md}{M_{\rm d}}
\newcommand{\Td}{T_{\rm d}}
\newcommand{\kd}{k_{\rm d}}
\newcommand{\kps}{\,\textstyle\rm{km~s}^{-1}}
\newcommand{\Mpc}{\,\textstyle\rm{Mpc}}
\newcommand{\coa}{CO(1$\rightarrow$0) }

\newcommand{\msun}{\,{\rm M}_{\odot}}
\newcommand{\jy}{\,\textstyle\rm{Jy}}
\newcommand{\K}{\,\textstyle\sc{k}}

\lefthead{Frayer et al.}  
\righthead{Submillimeter Imaging of VV\,114}

\begin{document}

\title{Submillimeter Imaging of the Luminous Infrared Galaxy Pair VV\,114}

\author{D.\ T.\ Frayer\altaffilmark{1}, 
R.\ J.\ Ivison\altaffilmark{2},
I.\ Smail\altaffilmark{3},
M.\ S.\ Yun\altaffilmark{4}, and
L.\ Armus\altaffilmark{5}}

\altaffiltext{1}{Astronomy Department, California Institute of Technology
105--24, Pasadena, CA  91125, USA} 

\altaffiltext{2}{Department of Physics and Astronomy, University College
London, Gower Street, London, WC1E 6BT, UK}

\altaffiltext{3}{Department of Physics, University of Durham, South Road,
Durham, DH1 3LE, UK}

\altaffiltext{4}{National Radio Astronomy Observatory, P.O.\ Box 0,
Socorro, NM  87801, USA}

\altaffiltext{5}{SIRTF Science Center, California
Institute of Technology 100--22, Pasadena, CA 91125, USA}

\begin{abstract}
 
We report on 450- and 850-$\micron$ observations of the interacting
galaxy pair, VV\,114\,E\,+\,W (IC\,1623), taken with the SCUBA camera on
the James Clerk Maxwell Telescope, and near-infrared observations taken
with UFTI on the UK Infrared Telescope.  The system VV\,114 is in an
early stage of a gas--rich merger.  We detect submillimeter (sub-mm)
emission extended over $30\arcsec$ (12 kpc) and find a good correlation
between the spatial distribution of the sub-mm and CO emission.  Both
the CO and sub-mm emission peak near the reddest region of VV\,114\,E
and extend toward VV114\,W.  The bulk of the sub-mm emission resides in
the central region showing the largest CO velocity gradients, which is
thought to mark the kinematic centroid of the merger remnant.  We
derived a total dust mass of $1.2\times 10^{8} \msun$, assuming a
power-law distribution of dust temperatures.  The sub-mm observations
suggest that the majority of the dust is relatively cool ($\Td \sim$
20--25\,K), and the total dust mass is about 4 times higher than that
inferred from the {\em IRAS} data alone. The system will likely evolve
into a compact starburst similar to Arp\,220.

\end{abstract}


\section{INTRODUCTION}

Many extremely luminous infrared starbursts ($L_{\rm IR} > 10^{11}
L_{\sun}$) are merging systems, containing large reservoirs of molecular
gas and dust (Sanders \& Mirabel 1996, and references therein).  To
understand the evolution of the ISM and star formation in these systems,
we require imaging observations of the molecular gas and dust reservoirs
associated with the star-forming regions.  Much of what is known about
these systems has been inferred from interferometric CO observations
(\eg Scoville, Yun, \& Bryant 1997; Downes \& Solomon 1998).  Our
knowledge of the distribution of thermal dust emission in merging
systems is lacking, due to the limited spatial resolution of the {\em
Infrared Astronomical Satellite} ({\em IRAS}\,).  With the commissioning
of the new Sub-mm Common-User Bolometer Array (SCUBA, Holland et al.\
1999) camera on the James Clerk Maxwell Telescope (JCMT), we can now
directly image the thermal dust emission at relatively high ($\sim
8-15\arcsec$) spatial resolution.

The interacting galaxy pair VV\,114 is only 80 Mpc away ($H_o=75\kps
\Mpc^{-1}$) and is comprised of two stellar components separated by
$15\arcsec$ (6 kpc in projection), designated VV\,114\,E and VV\,114\,W
(Knop et al.\ 1994).  The extreme infrared colors of VV\,114\,E indicate
the presence of a large concentration of dust, while VV\,114\,W is
relatively unobscured at optical wavelengths (Knop et al.\ 1994).  The
CO emission is distributed along a bar-like structure located in between
the optical components (Yun, Scoville, \& Knop 1994).  The extended
nature of the CO emission as well as the non-circular gas kinematics
suggests that VV\,114 is in an early stage of a gas-rich merger (Yun et
al.\ 1994).  The gas in VV\,114 has already begun to flow into its
central regions, while the stellar components are still well separated.
This is consistent with theoretical models of young mergers (Barnes \&
Hernquist 1991, 1996).

In addition to the capability of SCUBA to resolve the VV\,114 system,
the sub-mm observations at 450 and 850\,$\mu$m allow us to constrain
the amount of cool material present ($\Td \la 30 \K$).  Preliminary
evidence for the existence of cold dust in galaxies comes from the
gas-to-dust ratios estimated using the warm dust masses derived from
{\em IRAS}.  The typical gas--to--dust ratios found for the bright
{\em IRAS} galaxies are about an order of magnitude larger than the
Galactic value derived from extinction studies, which includes dust at
all temperatures (Devereux \& Young 1990).  A simple explanation for
this discrepancy is that the majority of the dust in bright {\em IRAS}
galaxies is cooler ($\Td \la 30 \K$) than the warm material observed
by {\em IRAS}.  Observations with SCUBA can directly search for this
component and provide a test of this hypothesis.

\section{OBSERVATIONS}

\subsection{Sub-mm Imaging}

Fully-sampled images of VV\,114 were obtained using the SCUBA sub-mm
camera during 1997 July and 1998 May.  We used the 91-element short-wave
array at 450\,$\micron$ ($7\farcs8$ FWHM beam) and the 37-element
long-wave array at 850\,$\micron$ ($14\farcs7$ FWHM beam) during
excellent conditions.  Typical 850-$\micron$ zenith opacities were
approximately 0.2, determined from hourly skydips.  We obtained 7.7\,ks
of useful on-source integration time at 850\,$\micron$ and 5.1\,ks at
450\,$\micron$.

Fluxes were calibrated against Uranus and the flux densities determined
for VV\,114 at 450 and 850\,$\micron$ are accurate to $\pm 18$\% and
$\pm 14$\%, respectively, including errors induced by opacity
uncertainties and the absolute flux uncertainty of Uranus.  The
uncertainty in the 450$\micron$/850$\micron$ flux ratio is approximately
$\pm7$\%, which represents the full dispersion in the measurements made
on the separate nights.  The error in the 450$\micron$/850$\micron$ flux
ratio is less than total uncertainty at either wavelength since the
absolute flux calibration error of Uranus cancels for the ratio.
Pointing was checked regularly using a nearby blazar.  We estimate the
overall positional accuracy of the submillimeter maps to be $\pm
2\arcsec$.

\subsection{Near-infrared Imaging}

Near-infrared, $J$- and $K$-band, imaging of VV\,114 was obtained during
the night of 1998 October 19 using the new UKIRT Fast-Track Imager
(UFTI) on the 3.8-m UK Infrared Telescope (UKIRT) on Mauna Kea.  UFTI
comprises a $1024^2$ Hg\,Cd\,Te HAWAII array cooled to $77\K$ and
sensitive from 0.8 to 2.5\,$\micron$. The pixel scale is $0\farcs091$,
allowing Nyquist sampling of the best seeing experienced on UKIRT, and
the field of view is $92\arcsec$.  The observations of VV\,114 consisted
of three sets of 9 exposures, two in $K$ ($2.2\micron$) and one in $J$
($1.2\micron$), each of 30\,s, dithered on a $3\times 3$ grid with
$9\farcs1$ spacing.  These were interspersed by similar length
observations of an offset sky region and dark exposures to allow the
removal of the sky and instrumental backgrounds.  The final stacked
exposures represents 270\,s in $J$ and 540\,s in $K$.  The tip/tilt
secondary on UKIRT was used to provide standard adaptive correction
within our field, although the conditions were non-photometric and the
seeing achieved was only mediocre, $\sim 0\farcs7$.  Nevertheless,
these images represent a substantial improvement in resolution over
those presented by Knop et al.\ (1994).

\section{Results}

Figures~1a \& 1b show the 450- and 850-$\micron$ images of VV\,114 taken
with SCUBA.  The integrated flux densities summed over the emission
regions are $2.43\pm0.44\jy$ and $0.273\pm0.038\jy$ at 450$\micron$ and
850$\micron$, respectively (Table~1).  At 850$\micron$, we detect
emission extended in the east-west direction, while at 450$\micron$ we
appear to resolve the central emission into two peaks (450$\micron$[1]
and 450$\micron$[2] in Table~1).  Given that the separation of the two
450$\micron$ peaks is near the resolving limit of SCUBA, observations at
higher resolution are required to confirm this double--peaked
morphology.  When the 450-$\micron$ data are convolved to the resolution
of the 850-$\micron$ map, the data at both wavelengths show a similar
shape, size, and position.  At this resolution ($14\farcs7$), we fail to
detect any significant variations in the $S(450\micron)/S(850\micron)$
flux ratio.  Integrating over all emission regions, we find a flux ratio
of $8.9\pm0.6$, which is also the ratio measured for the peak of the
sub-mm emission (at a resolution of $14\farcs7$).  The consistencies in
the 450-$\micron$ and 850-$\micron$ data would be expected if the bulk
of the sub-mm emission is on the black--body tail of spectral energy
distribution with dust temperatures $T\ga 10\K$.  Observations at higher
resolution are required to search for possible variations in the
$S(450\micron)/S(850\micron)$ ratio on smaller spatial scales.

The sub-mm emission lies along the CO bar-like structure (Fig.~1c),
extending east-west between the two optical components.  Both the CO and
sub-mm emission peak near the brightest $K-$band source of VV\,114 which
is by far the reddest part of the VV\,114 system, as indicated by the
$J-K$ color map (Fig.~1d).  Although the NIR data were not taken in
photometric conditions, an approximate $J-K$ magnitude scale was derived
by matching the data with the previously determined $J-K$ color of
VV\,114\,W (Knop et al.\ 1994).  The higher--resolution data presented
here suggests a color of $(J-K)\simeq 3$ for the bright compact red
component in VV\,114\,E.  These results are consistent with those found
for the Near-Infrared Camera and Multi-Object Spectrometer (NICMOS) data
taken with the Hubble Space Telescope (Scoville et al.\ 1999).

\section{DISCUSSION}

\subsection{Comparison of the Sub-mm Maps with Data at Other Wavelengths}

The sub-mm, radio continuum, and CO emission from VV\,114 all have
roughly the same spatial extent.  However, there are significant
differences in their detailed structure.  In particular, the radio
sources (Condon et al.\ 1990; Condon et al.\ 1991) are located near the
near-infrared (NIR) emission peaks (Knop et al.\ 1994), suggesting that
these regions are responsible for the majority of the ongoing star
formation.  The sub-mm and CO peaks do not appear to be correlated with
the radio and NIR peaks but tend to lie primary in between
VV\,114\,E\,+\,W.

The CO peak lies approximately $2\arcsec$ from the submillimeter peak,
which is consistent within the positional uncertainties of the data
sets.  The general similarities of the CO and sub-mm maps suggest that
both the \coa and sub-mm emission probe the same material: the molecular
gas and dust reservoir associated with the merger event.  Despite these
consistencies, there are differences in their detailed structure which
may indicate optical depth or excitation effects in the CO emission.
The CO emission is distributed more smoothly along the bar-like
structure between VV\,114\,E and VV\,114\,W, while the sub-mm emission
is more strongly peaked near VV\,114\,E.  The tidal tails are also more
apparent in the CO map, while the 450$\micron$ emission is more extended
in the east and south-west directions.  The excess sub-mm emission in
the east and south-west directions may be associated with {\sc Hi} tidal
debris infalling into the central regions (Hibbard \& Yun 1999).

VV\,114\,E is located near a peak of the sub-mm emission which is
consistent with the high level of dust obscuration inferred from its
extremely red NIR colors (Fig.~1d; also see Knop et al.\ 1994).
VV\,114\,W, on the other hand, lies along a line of sight which is
nearly free of sub-mm emission (Fig.~1a) and is consistent with the low
level of extinction estimated at NIR wavelengths (Knop et al.\ 1994).
The NIR data show that the star formation is occurring globally
throughout the system, but the most active site is in the nucleus of
VV\,114\,E, which is associated with the brightest radio sources in the
8.4-GHz sub-arcsec resolution map (Condon et al.\ 1991).  The ISO
15-$\micron$ continuum also peaks strongly on the eastern nucleus of
VV\,114 (Yun et al. 1999).  VV\,114\,E, itself, is comprised of two
bright nuclear sources in the NIR (Knop et al. 1994).  The north-east
component of VV\,114\,E is strongest in $J$ and is associated with the
brightest compact radio source (marked by a triangle symbol in Fig. 1)
in the high-resolution radio map (Condon et al. 1991).  The red
south-west component of VV\,114\,E is strongest in $K$ (marked by a
$\times$ symbol in Fig. 1) and is associated with a region containing
several bright, compact radio sources.  This red south--west component
of VV\,114\,E is nearest to the position of the peaks in the sub--mm and
CO emission, but is still offset by $3\farcs5$--5$\arcsec$.  This
positional offset is larger than the positional errors in the data sets
and hence may be significant.

Although most of the current star formation is occurring in and around
the stellar components seen in the NIR (Knop et al.\ 1994), the central
sub-mm and CO emission regions between the optical components may mark
the location of a future major starburst given that these regions show
the largest CO velocity gradients (Yun et al.\ 1994) and contain the
bulk of the molecular gas and dust in the system.  The apparent
displacement of the sub-mm and CO emission from the most active regions
of star-formation is consistent with the low HCN/CO ratio observed in
VV\,114 (Gao 1996).  Typical infrared luminous galaxies show enhanced
HCN/CO ratios, presumably due to the high density of the molecular gas
in the starburst regions (Gao 1996).  The low HCN/CO ratio in VV\,114
may indicate that the bulk of the molecular gas is in regions of lower
density and is displaced from the most active star--forming regions.
Although the sub-mm emission of VV\,114 is currently extended over
12~kpc, it is likely to evolve into a more compact starburst system
similar to that seen for Arp~220 (Downes \& Solomon 1998; Sakamoto et
al.\ 1998).  In the context of the numerical models for merging galaxies
(Barnes \& Hernquist 1996), VV\,114 is an early-stage merger where the
gas is accumulating within the central regions in advance of the stars.
The system has yet to trigger a strong compact starburst at the
dynamical centroid as seen for Arp~220 and other compact ultraluminous
galaxies (Downes \& Solomon 1998).

\subsection{Estimating the Total Dust Mass}

Observations at submillimeter wavelengths provide a better estimate of
the total dust mass than that inferred from the {\em IRAS} data alone,
since {\em IRAS} was not sensitive to cool dust.  Assuming optically
thin dust emission, the mass in dust is given by $\Md =
S_{\nu}\,D^{2}\,(\kd\,B[\nu,\Td])^{-1}$, where $B(\nu,\Td)$ is the
black-body function for a frequency $\nu$ and a temperature $\Td$, $D$
is the distance, and $\kd$ is the dust absorption coefficient.  We adopt
$\kd = 10$ g$^{-1}$ cm$^2\,(\lambda/250\micron)^{-\beta}$, where
$\beta=1$ for $\lambda < 250 \micron$ (Hildebrand 1983).  At longer
wavelengths ($\lambda > 250 \micron$), the typical values for $\beta$
range between 1 and 2.  Empirically at long wavelengths, the Galactic
cirrus is well fitted by $\beta\simeq 1.5$ (Masi et al.\ 1995), while
Galactic star-forming regions have $\beta\simeq 2$ (Chini et al.\ 1986;
Lis et al.\ 1998).

From the submillimeter measurements, we can directly estimate the value
of $\beta$ appropriate for VV\,114 as a function of temperature.  The
integrated sub-mm flux density ratio is $S(450\micron)/S(850\micron) =
8.9 \pm 0.6$.  The value for $\beta$ is calculated from
$S(450\micron)/S(850\micron) = (450/850)^{-\beta} \times B(666\,{\rm
GHz},\Td)/B(353\,{\rm GHz},\Td)$.  At $\Td = 20\K$, the data imply
$\beta=2.15 \pm 0.11$.  At the temperature of $\Td = 41$ inferred from
the {\em IRAS} data (Soifer et al.\ 1989), the sub-mm data imply
$\beta=1.76 \pm 0.11$.  Since the mass--weighted average temperature of
the dust is expected to be less than or equal to the {\em IRAS} dust
temperature, the lower limit for $\beta$ in VV\,114 is $\beta \geq
1.65$.  In fact, we find (see below) that most of the dust is relatively
cool in VV\,114 ($T\simeq$ 20--25$\K$) suggesting that $\beta \simeq 2$,
which is similar to that observed for Galactic star-forming regions
(Chini et al.\ 1986).

We fit thermal dust spectral energy distributions (SEDs) to the
observational data in order to estimate the total dust mass in VV\,114.
Figure~2a shows models appropriate for a temperature of $\Td = 41\pm4
\K$ derived from the {\em IRAS} $60\micron/100\micron$ flux density
ratio (Soifer et al.\ 1989).  The dust mass inferred for this
temperature is $(3\pm\stackrel{2}{{_1}})\times10^{7}\msun$
(Table~2). The uncertainty in $\kd$ provides an additional uncertainty
of about a factor of two in the dust mass.  None of the single-component
temperature models fit the sub-mm data.  The low $\beta$ models are
inconsistent with the $S(450\micron)/S(850\micron)$ flux density ratio,
and for $\beta\simeq 2$ the sub-mm fluxes are significantly higher than
those expected from the {\em IRAS} data.  A simple interpretation of the
data is the presence of cooler material which was not detected by {\em
IRAS}.  Figure~2b shows a three-component fit to the data, assuming
$\beta=2$ at $\lambda > 250\micron$.  The mass of dust residing in the
cool component ($\Td = 20\K$) is 1--$2\times 10^{8}\msun$, which is
significantly larger than that found in the warm {\em IRAS} component
($\Td = 41\K$, $0.3\times 10^{8}\msun$).  By varying the allowed dust
temperatures, the relative amount of dust in the warm (37--45 $\K$) and
cool ($\sim$15--25 $\K$) components can be modified by factors of 2--3.
The amount of dust in the hot component ($\Td = 150\K$), constrained by
the 12 and 25$\micron$ data, is negligible (Table~2).

More complicated models, which include several different grain types,
cool cirrus, and warm starburst components could also be fitted to the
data (Rowan-Robinson 1992; Andreani \& Franceschini 1996).  Although
such models may be more realistic, the number of model parameters
exceeds the number of observational data points for VV\,114.  For the
model with three temperature components (Fig.~2b), there are six free
parameters (mass and temperature of each component) for only six data
points.  In reality, there is likely to be a continuous range of dust
temperatures.

A simple approach is to adopt a power-law distribution of dust
temperatures; $dM_{\rm d}/dT \propto T^{-x}$ (Xie, Goldsmith, \& Zhou
1991).  The total dust mass is determined by integrating the
distribution function over a range of temperatures from $T_{\rm low}$ to
$T_{\rm high}$.  The upper limit to the dust temperature has little
effect on the total dust mass (\eg de Muizon \& Rouan 1985).  We assume
$T_{\rm high} = 200\K$.  Fig.~2c shows dust SEDs computed for several
different temperature distributions.  Considering the small number of
free parameters ($x$, $T_{\rm low}$, and $\Md$[total]), the shape of the
computed SEDs match the observational data remarkably well. The best fit
to the 60--850$\micron$ observational data occur for $T_{\rm low} =22$
and $x \simeq 6$.  The 12--25 $\micron$ data are not fitted well with a
single power-law distribution, but these data have little effect on the
total dust mass.  Since $\Md \propto \Td^{-5}$ for dust in thermal
equilibrium (Soifer et al.\ 1989), we expect $x \simeq 6$.  With this
piece of theoretical insight, the SED is effectively fitted by only
$T_{\rm low}$ and $\Md$(total), after assuming an appropriate form for
$\kd(\lambda)$.  The best fit ($x = 6$; $T_{\rm low} = 22\K$) implies a
total dust mass of $1.2 \times 10^{8}\msun$.  Assuming a power-law
distribution of dust temperatures, the total dust mass is well
constrained by the data.  The error in fitting the dust mass is only
about 25\%.  However, considering the uncertainty in $\kd$ at sub-mm
wavelengths (Hughes, Dunlop, \& Rawlings 1997), the derived dust mass is
uncertain by approximately a factor of two.

\subsection{Implications for Cool Dust in Luminous {\em IRAS} Galaxies}

Most of the dust in VV\,114 is at a temperature of $\Td =$20--25$\K$.
Based on the SCUBA data, the total dust mass derived for VV\,114 is four
times larger than that derived from the 60--100$\micron$ {\em IRAS}
data.  Since the {\em IRAS} colors for VV\,114 are typical for its
infrared luminosity (Soifer \& Neugebauer 1991), we could expect to find
similar results for other luminous {\em IRAS} sources.  The existence of
cool dust ($\Td < 30\K$) undetected by {\em IRAS} is not a surprising
result.  As stated in \S1, the high gas-to-dust ratios in galaxies
derived from the {\em IRAS} data of spiral galaxies imply the presence
of cool material (Devereux \& Young 1990).  Millimeter observations of
spiral galaxies also suggest massive cool dust reservoirs with $\Td \sim
10$--$20\K$ (Gu\'{e}lin et al.\ 1993, 1995; Franceschini
\& Andreani 1995).  Recent SCUBA observations of NGC\,891
support these results with the detection of a cool dust reservoir which
is over an order of magnitude more massive than its warm dust component
(Alton et al.\ 1998).

The results for VV\,114 indicate that even very luminous {\em IRAS}
galaxies can have cool dust reservoirs.  Based solely on the {\em IRAS}
data, the gas-to-dust ratio for VV\,114 is $M(\hh)/\Md = 1100$ (Sanders,
Scoville, \& Soifer 1991), which is much higher than the Galactic value
of about 100 (Devereux \& Young 1990).  By using the dust mass implied
by the sub-mm data, we find a more realistic value of $M(\hh)/\Md \simeq
300$.  These gas--to--dust ratios assume the Galactic CO to H$_2$
conversion factor.  If the Galactic value is not applicable for
ultraluminous galaxies (Solomon et al.\ 1997), then the gas--to--dust
ratio for VV\,114 could be reduced further by a factor of 2--4.
Observations of other luminous {\em IRAS} galaxies will test the
generality of these results.

The cool dust temperature for VV\,114 may be related to its early-merger
evolutionary phase.  More evolved, luminous mergers have warm dust
emission associated with a compact starburst and/or an active galactic
nucleus (Mazzarella, Bothun, \& Boroson 1991).  For example, the sub-mm
data for the evolved, compact starburst, Arp~220 (Rigopoulou, Lawrence,
\& Rowan-Robinson 1996), show no evidence for excess sub-mm emission
associated with cool dust and can be fitted with a single-temperature,
warm dust model of $\Td = 47\K$ (Klaas et al.\ 1997).  The dust may be
cooler in VV\,114 because the majority of the dust is displaced from the
star-forming regions traced by the NIR and radio emission peaks.  Given
the radio-to-FIR relationship (Helou, Soifer, \& Rowan-Robinson 1985),
we expect to find warm dust within the star-forming regions, especially
near the most active region, VV\,114\,E.  As VV\,114 evolves into a
compact gaseous system similar to Arp~220, the star formation is
expected to increase in the central regions and the dust temperature
should rise accordingly.

\section{CONCLUSIONS}

We present sub-mm maps of the young merger system VV\,114.  We detect
sub-mm emission in excess of that expected from the {\em IRAS} data.
This sub-mm excess suggests the presence of a cool, massive component of
dust.  By fitting a variety of dust models to the SED of VV\,114, we
derive a total dust mass of approximately $1\times10^{8}\msun$, a
temperature of $\Td \simeq$ 20--25$\K$, and $\beta\simeq 2$.  The
majority of the dust is located in between the optical components,
VV\,114\,E + W, near the suspected dynamical center of the merger
remnant.  The sub-mm emission regions correlate well with the CO
emission regions, but do not correlate well with the peaks in the radio
and NIR emission regions which are thought to trace the star-formation
activity.  The fact that the sub-mm emission is displaced from the most
active regions of star formation may explain the cool temperature for
the majority of the dust.  Given the extremely large reservoir of gas
and dust available in the central regions, VV\,114 is expected to evolve
into a more luminous central starburst, similar to that seen for
Arp~220.

\acknowledgments

We thank the staff at the JCMT and UKIRT who have made these observations
possible, and the Director of the JAC, Ian Robson, for the provision of
discretionary time.  DTF acknowledges support from NSF grant AST
96--13717 to the Owens Valley Millimeter Array; RJI and IRS
acknowledge support from a PPARC Advanced Fellowship and a Royal
Society Fellowship, respectively.

\figcaption[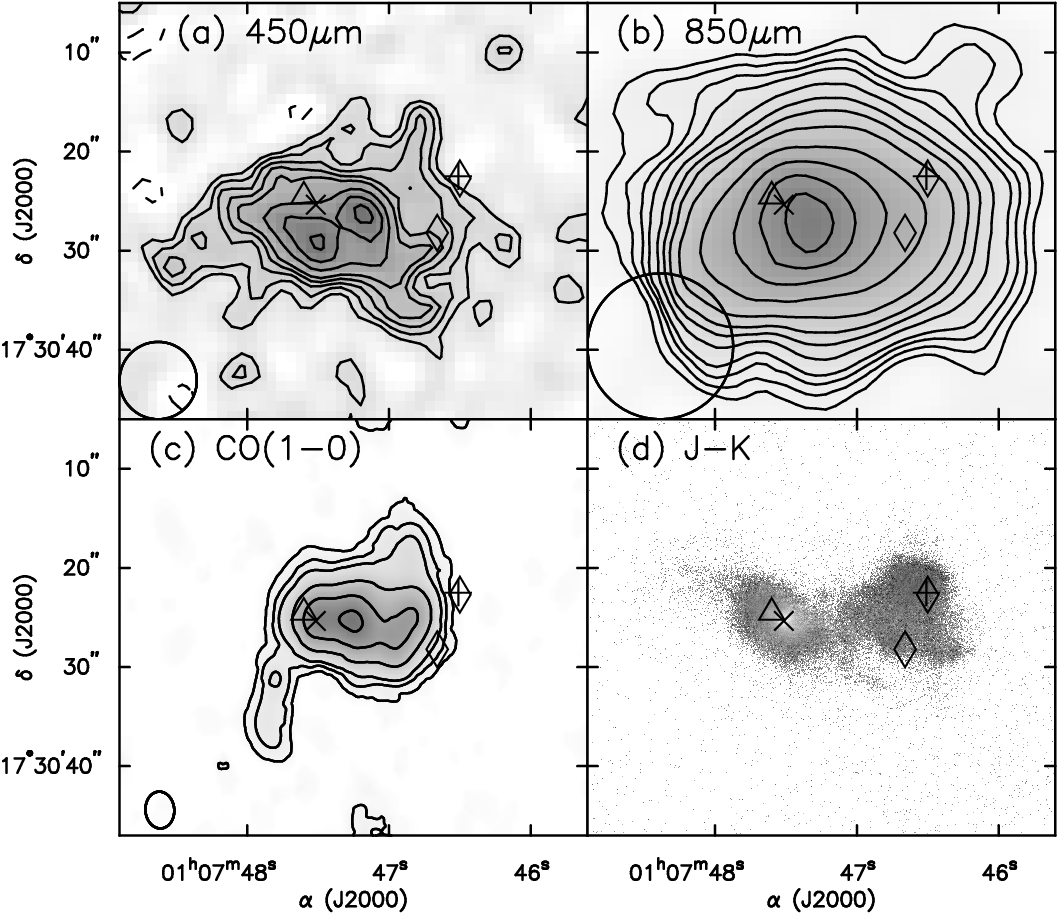]{SCUBA (a) 450-$\micron$ and (b) 850-$\micron$
maps.  The 1$\sigma$ rms levels are 75 and 4.4~mJy\,beam$^{-1}$ for the
450- and 850$\micron$ data, respectively.  Contour levels are plotted at
$\sigma\,\times$ ($-$2, 2, 3, 4, 5, 6, 8, 10, 12) for the 450-$\micron$
map and $\sigma\,\times$ (3, 4, 5, 6, 8, 10, 12, 15, 20, 25, 30, 35) for
the 850-$\micron$ map.  The 2$\micron$ NIR peaks, signifying the
positions of VV\,114\,E + W are shown by the $\times$ and $+$ symbols
(Knop et al.\ 1994).  The 8.4-GHz radio peak is shown by the $\triangle$
symbol (Condon et al.\ 1991) and the $\Diamond$ symbols show peaks in
the optical $R$-band emission (Knop et al.\ 1994).  Panel (c) shows the
integrated \coa emission (Yun et al.\ 1994).  Contour levels are plotted
at $1\jy\kps$\,beam$^{-1}$ $\,\times$ (5, 10, 15, 30, 50, 70).  For
panels (a), (b), and (c) the beam sizes are shown in the lower left of
each panel.  Panel (d) shows the NIR $(J-K)$ color map, where lighter
pixels indicate redder colors and darker pixels represent bluer colors.
The color map ranges from $(J-K) =0.9$ for the blue regions of
VV\,114\,W to the very red compact region in VV\,114\,E which has a
color of $(J-K) =3.0$ at the observed resolution ($0\farcs7$).}

\figcaption[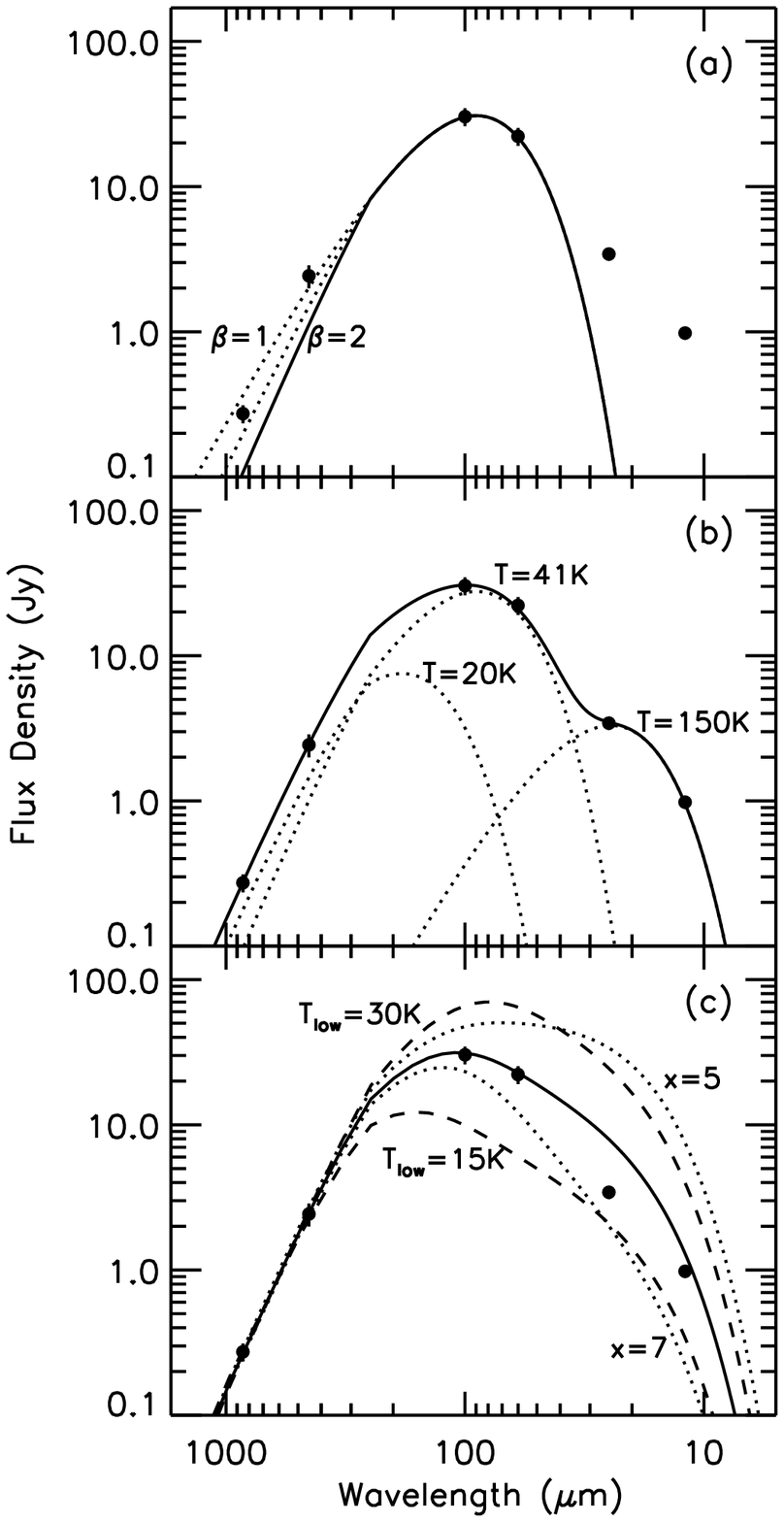]{Model fits to the spectral energy distribution of
VV\,114.  The {\em IRAS} data points are from Soifer et al.\ (1989).
Panel (a) shows single temperature models ($\Td = 41\K$) based on the
{\em IRAS} 60--100 $\micron$ data for $\beta=1,1.5$ (dotted lines),
and $\beta=2$ (solid line).  No single temperature model can fit the
data.  Panel (b) shows a three-component model fit to the data,
assuming $\beta=2$ for $\lambda>250\micron$.  The three different
components are plotted as dotted lines, and their sum is shown by the
solid line.  Panel (c) shows models based on a distribution of dust
temperatures; $dM_{\rm d}/dT \propto T^{-x}$ (see \S4.2).  The solid
line represents the best model fit with $T_{\rm low} = 22\K$ and $x =
6$.  The dotted lines show models with different values of $x$, while
the dashed lines show fits based on different values of $T_{\rm
low}$.}

\clearpage
\begin{deluxetable}{lccc}
\tablenum{1}
\tablecaption{Submillimeter Observational Results}
\tablewidth{300pt}
\tablehead{\colhead{Parameter}&\colhead{450$\micron$[1]}
&\colhead{450$\micron$[2]}&\colhead{850$\micron$}}
\startdata
$\alpha$(J2000)\tablenotemark{a,b}\hspace{1cm} $01^{\rm h}07^{\rm m}$&
$47\fs55$& $47\fs15$& $47\fs35$\nl
$\delta$(J2000)\tablenotemark{a,b}\hspace{1cm}$-17\arcdeg 30\arcmin$
&$28\farcs9$ &$26\farcs1$ &$27\farcs2$\nl
&&&\nl
Beam (FWHM) \dotfill\ &\multicolumn{2}{c}{$7\farcs8$} &$14\farcs7$\nl
Flux (Jy)\tablenotemark{c}\dotfill\ &\multicolumn{2}{c}{$2.43\pm0.44$}
&$0.273\pm0.038$\nl 
\enddata 
\tablenotetext{a}{Positional uncertainty is $\pm 2\farcs0$.}
\tablenotetext{b}{Derived from a Gaussian fit to the peak.}
\tablenotetext{c}{Total integrated flux.}
\end{deluxetable}

\begin{deluxetable}{lrrcc}
\tablenum{2}
\tablecaption{Dust Models}
\tablewidth{450pt}
\tablehead{\colhead{Model}&\colhead{Fitted Data}
&\colhead{$\Td(\K)$}& \colhead{$\Md (10^{7}\msun)$} &\colhead{Notes}}
\startdata
Single Component & 60--100$\micron$ & 41 & 3.0 & Fig.\ 2a \nl
&&&&\nl
Three Component & 12--850$\micron$ &  20 & 13   & Fig.\ 2b \nl
                &                  &  41 & 2.7  & Fig.\ 2b \nl
                &                  & 150 &0.0018& Fig.\ 2b \nl
&&&&\nl
Temperature Distribution& 450--850$\micron$ & 15 & 21 & $\Td=T_{\rm low}$,
$x=6$, Fig.\ 2c\nl
& 450--850$\micron$ & 30 & 8 & $\Td=T_{\rm low}$, $x=6$, Fig. 2c\nl
& 60--850$\micron$ & 22 & 12 & $\Td=T_{\rm low}$, $x=6$, Fig. 2c\nl
\enddata
\end{deluxetable}

\setcounter{figure}{0}

\newpage
\begin{figure}
\includegraphics{fig1.ps} \vspace*{8.in}
\caption{ }
\end{figure}

\newpage
\begin{figure}
\includegraphics{fig2.ps} \vspace*{8.in}
\caption{ }
\end{figure}

\end{document}